\documentclass[12pt]{article}
\usepackage{amsmath,epsfig}
\usepackage{amssymb}

\begin{document}
\section*{On a Generalized Two-Fluid
Hele-Shaw Flow}

\begin{center}
 V.M.Entov$^{*}$ P.Etingof$^{**}$
\vskip 12pt

$^{*}$ Institute for Problems in Mechanics of the Russian Academy
of Sciences, 101-1, prospekt Vernadskogo, 119526, Moscow, Russia,

entov@ipmnet.ru

\noindent $^{**}$ Department of Mathematics, 
Massachussets Institute of Technology,  Cambridge, USA

etingof@math.mit.edu

\end{center}\vskip 12pt

{\small Analysis of displacement in a Hele-Shaw cell and porous
media is a source of a multitude of mathematical problems which
provide some insight into general features of nonlinear boundary
dynamics(\cite{EEK},\cite{HOck},\cite{VarEt}). Here, we consider a
slightly modified version of the classical problem of flow in a
potential external field which displays some new features related
to existence of singularities of the external field. The study was
prompted by the interest in coupled flow phenomena in saturated
porous media in the presence of electric current (``electrokinetic
phenomena"). This specific case will be briefly discussed later.
However,  eventually it became clear that the topic deserves study
{\it per se}.
 Throughout the text we will use expressions `Hele-Shaw flow' and `flow through a porous medium' interchangeably as synonyms due to the
well known analogy between the Darcy law for a porous medium and
the flow  rule in a thin gap between two parallel walls.}

\section{The Problem Statement}
We assume that fluid occupies a finite domain $D(t)$ in the plane
$(x,y)$ of a Hele-Shaw cell of gap thickness $b$.
The flow velocity ${\bf w}$ is determined by the
flow rule
$$ {\bf w}=-{k\over \mu} \nabla p +{k\over \mu}{\bf g}. \eqno(1.1)$$
Here, $p$ is the fluid pressure, $\mu$ is the fluid viscosity,
$k=b^2/12$ is the gap "permeability", ${\bf g}=\{g_x,g_y\}$ is the
body force due to an external field. We assume the body force to
be potential,
$$ {\bf g}=-\nabla \Psi(x,y).  \eqno(1.2)$$
Two familiar examples are the gravity force and the centrifugal force, corresponding
to
$$ \Psi=\rho g h, \quad {\rm and}\quad \Psi= -{1\over 2}\rho \omega^2 r^2,   \eqno(1.3)$$
respectively. Here, $\rho$ is the fluid density, $h$ is the height
above a datum, $g$ is the acceleration due to gravity, $r$ is the
distance from the rotation axis normal to the cell plane; $\omega$
is the rotation rate. These cases allow thorough study; they are
considered in particular in \cite{EEK},\cite{EEK1}. Here, we are
going to study the case when the external potential $\Psi$ has
singularities within and/or outside the domain $D(t)$.

The flow field satisfies the continuity equation

$$ {\bf \nabla\cdot w}=\sum_{j=1}^Nq_j\delta(x-x_j,y-y_j). \eqno(1.4)$$

Here, $q_j$ are strengths (flow rates) of the point sources (sinks) within the
flow domain.
We assume that at the boundary $\Gamma=\partial D(t)$
the pressure vanishes,
$$ p(x,y)=0,\quad (x,y)\in \Gamma.
\eqno(1.5)$$
The boundary dynamics is governed by the relation
$$ v_n=w_n={\bf w\cdot n},
\eqno(1.6)$$
${\bf n}$ being the outward normal to $\Gamma$, and $v_n$ velocity of propagation of the  boundary in the normal direction.

 We assume now that the external potential field satisfies the
equation
$$ \Delta G=\sum_{m=1}^M Q_m\delta(x-x'_m,y-y'_m), \quad G=-{k\over\mu} \Psi,
\eqno(1.7)$$
in the entire plane $(x,y)$ with boundary condition
$$ |\nabla G| \rightarrow 0, \quad (x^2+y^2)\rightarrow
\infty.\eqno(1.8)$$
Therefore,
$$ G={\rm  Re} F(z),
\quad F=\sum_{m=1}^M {Q_m\over 2\pi}\ln (z-z'_m),$$
$$
z=x+iy; \quad z'_m=x'_m+iy'_m.\eqno(1.9)$$

Let us introduce the velocity potential
$$ \Phi(x,y)=-{k\over\mu}(p +\Psi).\eqno(1.10)$$
It satisfies the following problem:
$$ \Delta \Phi=-\sum_{j=1}^N q_j\delta(x-x_j,y-y_j), \quad (x_j,y_j)\in
D(t);
 \eqno(1.11)$$
$$ \Phi(x,y)=-{k\over\mu}\Psi(x,y)=G(x,y),\quad (x,y)\in \Gamma(t);
 \eqno(1.12)$$
$$v_n={\partial \Phi\over \partial n},\quad (x,y) \in \Gamma(t).
\eqno(1.13)$$
The last equation serves to describe the moving boundary dynamics.

The only difference with the usual Hele-Shaw problem is that
the flow potential does not vanish at the boundary, but should
be equal to a specified function of the boundary point.

Let now $u(x,y)$ be a harmonic function in a domain $D^*$ such,
that $D(t)$ remains within $D^*$. Then as a straightforward
generalization of the Richardson Theorem
\cite{Richardson1},\cite{Richardson2},\cite{EEK} we find:
$$ {d\over dt}\int_{D(t)}u dA
=\sum_{j=1}^N q_ju(z_j)+\int_{D(t)}\nabla G\cdot
\nabla u dA. \eqno(1.14)$$
The following chain of equalities proves statement (1.14):
$$ {d\over dt}\int_{D(t)}u dA=\int_{\partial D(t)}u{\partial \Phi
\over \partial n} dl=$$
$$
 \int_{\partial D(t)}
\Phi{\partial u\over \partial n} dl+
\int_{ D(t)}\left(u\Delta \Phi-\Phi\Delta u\right) dA=
$$
$$  \int_{\partial D(t)} G{\partial u\over \partial n} dl+
\sum_{j=1}^N u(z_j)q_j=\int_{D(t)}\nabla \cdot(G \nabla u) dA+
\sum_{j=1}^N u(z_j)q_j
$$
$$
=\sum_{j=1}^N q_ju(z_j)+\int_{D(t)}\nabla G\cdot
\nabla u dA.$$
The l.h.s. of Eq.(1.14) is a time derivative of a harmonic moment of
the domain $D(t)$.

This equation leads to explicit analytic techniques of predicting
domain evolution  provided that the operator
$\nabla G\cdot \nabla$ maps harmonic
functions to harmonic ones (and the domain
initially  belongs to a certain class of domains). It can be shown,
that it is possible only, if
$$ G=ax+by+ c(x^2+y^2)+d,\eqno(1.15)$$
with constant $a,b,c,d$. Essentially, it is a combination of
uniform (``gravity'') and axisymmetrical (``centrifugal force'')
fields treated previously \cite{EEK1},\cite{EEK}. Of course, for
$c\ne 0$, it can be reduced to pure rotation about a shifted axis.

Here, we will be interested primarily in equilibrium shapes of the
flow domain under combined action of the flow and the external
potential. In this case, one can derive effective solutions for
a much wider class of the external potentials.

\section{Steady-state shapes}


For the steady state (equilibrium) domain $D$ we arrive at the
following moment
 problem:
$$ \forall   u:\quad  \Delta u=0\quad {\rm in}\quad D,\eqno (2.1)$$
$$ \int_D\nabla G\cdot \nabla u dA=-\sum_{j=1}^N q_ju(z_j).\eqno (2.2)$$
(Of course, equilibrium domain can exist only if the net fluid flux vanishes,
 $\sum_{j=1}^N  q_j=0$.)
Now we assume that the external field has the form
$$ G(z)=\sum_{m=1}^M {Q_m\over 2\pi}\ln |z-z'_m|. \eqno (2.3)$$
In other words, it can be considered as an electric field
generated by a finite array of point charges in a plane. For
brevity sake, we will refer to it as an ``electric potential"
field. Note that $z'_m$ can be both inside and outside $D$. Let us
now introduce the corresponding complex potential $F(z)$ and `complex
current' $\omega(z)$:
$$ F(z)=G(z)+i\Psi(z)=\sum_{m=1}^M {Q_m\over 2\pi}\ln (z-z'_m);  \eqno (2.4)$$
$$ \omega(z)=F'(z)={\partial G\over \partial x}-i{\partial G\over \partial y}
=\sum_{m=1}^M {Q_m\over 2\pi (z-z'_m)}. \eqno (2.5)$$ (The complex
potential $F(z)$ is, generally speaking, multivalued, unless all
the `electric sources' are outside $D$.)
 Then  the moment equation (2.2)
can be written as
$$ J_D=\int_D\overline{\omega(z)}U'(z)dA
=-\sum_{i=1}^N q_iU(z_i),\eqno (2.6)$$
for an analytic function $U$ on a neighbourhood of $D$.

Integral $J_D$  in the l.h.s. of Eq.(2.6) converges
even if some $z'_m$ belong to $D$ as $\omega(z)$ has
simple poles at these points.
We write the integral as
$$ J_D=\int_D\overline{ \omega(z)}U'(z) dxdy={1\over 2i}\oint_{\partial D} \overline{\omega(z)}
U(z) d\overline z +\sum_{m:\ z'_m\in D}
{Q_m\over 2} U(z'_m).
\eqno (2.7)$$
The last transformation follows from the Green
Theorem; the sum in the r.h.s. accounts for contributions
of poles of $\omega$ in $D$.

Let us now choose
$$ U(z)={1\over \pi (w-z)}, $$
$w$ being a point outside $D$. Then Eq.(2.6) becomes
$$ {1\over 2\pi i}\oint_{\partial D}
{\overline{\omega(z) dz}\over w - z}=
\sum_{i=1}^N
{q_i\over \pi(w-z_i)}
-\sum_{m:\ z'_m\in D}
{Q_m\over 2\pi} {1\over w-z'_m}.
\eqno (2.8)$$
We denote the primitive of the function in
the r.h.s. of Eq.(2.8) by $h(w)$:
$$ h(w)= \sum_{i=1}^N
{q_i\over \pi}\ln(w-z_i) -\sum_{m:\ z'_m\in D} {Q_m\over 2\pi} \ln
(w-z'_m). \eqno (2.9)$$ The following Theorem due to Richardson
\cite{Richardson1} plays a crucial role in solving the problem of
finding the domain $D$:

{\it Theorem}

 Let $f: K\rightarrow D$ be a conformal
mapping that maps unit disk of the $\zeta$-plane onto D. Then the
function
$$ {d\over d\zeta}\left(\overline{F(f({1\over\overline {\zeta}}))}-
h(f(\zeta))\right)\eqno (2.10)$$
initially defined in a vicinity of the unit circle
extends analytically to a holomorphic function
in $K$.
\vskip 12pt
{\it Proof}. On the unit circle $\zeta=e^{i\varphi}$,
$\zeta=1/{\overline \zeta}$ and thus it suffices to show that the
differential $d(\overline{F(z)}-h(z))$ on $\partial
D$ extends to a holomorphic differential in $D$.

So according to the Cauchy Theorem it is necessary to check that
$$ \oint_{\partial D}{d(\overline{ F(z)}- h(z))\over t-z}=0,\quad
{\rm for} \quad t \notin D. \eqno (2.11)$$ However, it follows
directly from Eqs.(2.8),(2.9)
 and the fact that $dh$ is
holomorphic outside $D$.

This implies the following important corollary:
\vskip 12pt
{\it Corollary.}

The function ${d\over d\zeta} F(f(\zeta))$ is rational.
\vskip 12pt
{\it Proof}. For any function $f(z)$
denote
$$ f^*(z)=\overline{f(\overline z)}. $$
It implies immediately
$$ \overline{F(f(z))}=F^*(f^*(\overline {z})). $$

Then according to the Theorem (2.10) the differential
$$ d(F^*(f^*(1/ {\zeta}))-h(f(\zeta)))$$
is analytic in the unit disc.

 Therefore, the differential
  $${\frak D}(\zeta):= d(F^*(f^*({1\over \zeta})) $$
has the same singularities as $dh(f(\zeta))$ in the unit disk $K$,
i.e. simple poles at the
points
$$\zeta=f^{-1}(z'_m),\quad  z'_m\in D \quad {\rm and}\quad \zeta=f^{-1}(z_j),\quad
j=1,\ldots,N. $$
Therefore $dF(f(\zeta))$ has poles at the points
$$1/\overline{f^{-1}(z_j)},\quad  1/\overline{f^{-1}(z'_m)},\quad z'_m\in D$$
outside the unit disk.
On the other hand,
$dF(f(\zeta))$ has poles at the points
$$\zeta=f^{-1}(z'_m),\quad \quad z'_m\in D$$
within the unit disk.

Therefore, ${\frak D}(\zeta)$ has a finite number of poles, and
hence it is rational. Then determination of the precise form of
$f(\zeta)$ can be reduced to a set of nonlinear algebraic
equations.

\vskip 12pt Note, that this analysis can be in a standard way
extended on the limiting case of coalescence of hydrodynamic
sources and sinks corresponding to multipoles. In such a case, the
terms
$$ {q_j\over \pi (z-z_j)} $$
have to be replaced with the terms
$$ {\mu_j\over \pi (z-z_j)^n},\ n>1, \quad {\rm etc}. $$

\section{Harmonic potential}

\subsection{ Univalent $F(z)$}

Assume that $F'(z)\ne 0, \infty$, and that $F(z)$ is
 univalent in $D$. Then we can solve the problem in a more straightforward way.
We just notice that for the domain $\tilde D=F(D)$ one has
$$ \int_{\tilde D}U'(\tilde z)d\tilde x d\tilde y =-\sum q_jU(F_j);
\quad \tilde z=F(z);\quad F_j=F(z_j).\eqno(3.1)$$ However, this is
exactly the form of the moment equation that corresponds to
uniform external field with $G(\tilde z) =\tilde x$ (the Hele-Shaw
flow in the presence of gravity) \cite{EEK}.    If we take $u(z)=z^n$,
Eq.(3.1) becomes
$$\tilde M_{n-1}\equiv \int_{\tilde D}{\tilde z}^{n-1}d\tilde x d \tilde y
=-\sum q_j F_j^n/n,\eqno(3.2) $$ so that all moments are specified
at given $F(z)$.

Letting $n=1$ we have

$$ \tilde S=-\sum q_j F_j>0.\eqno(3.3) $$

It is a necessary condition for the existence of a steady-state solution.
In particular,
if the flow is generated by a dipole, (a
source-sink doublet of strength $\pm q=\pm \mu/(2\epsilon)$ at
$z=\pm \epsilon$)
then as $\epsilon \rightarrow 0$ the r.h.s. of Eq.(3.3) tends to
$\mu F'(0)$, and Eqs.(3.2) becomes
$$ M_0\equiv \tilde S= \int_{\tilde D} d\tilde x d \tilde y=
\mu;\quad M_n=0, \ n=1,2,3,\ldots \eqno(3.4) $$ Obviously, in the
$F$-plane the equilibrium domain is a circle of the radius
$$ R_0= \sqrt{\mu/\pi}.$$
Thus,  there is a fixed value of the area of the equilibrium
domain in the potential plane for which such a domain exists.
Obviously, the equilibrium in this case is due to a fine balance
between hydrodynamic and external forces, and is unstable.

The above elementary example is generic in the sense that for given
set of hydrodynamic and electric sources the area of the equilibrium domain,
provided such a domain  exists,
can assume only a discrete set of values, if the electric sources are outside
 the domain, so that $F(z)$ is analytic in $D$.

It is reasonable to ask about the fate of a domain that evolves under
 combined action of the balanced hydrodynamic sources  and the external
 field starting from a non-equilibrium shape. While in general case the
 answer is beyond our capacities, some insight can be derived from the
 simple case of ``gravity", i.e. uniform potential field,
$$ F(z)=g\rho z.\eqno(3.5)$$
Then the moments dynamics equation (1.14) becomes
$$ {d\over dt} \int_{D(t)} u dA=\sum_{i=1}^N q_iu(z_i)-C \int_{D(t)} {\partial
u\over\partial x} dA; \quad C={\rho g k\over \mu}.\eqno(3.6)$$ If
we take now
$$ u(z)={1\over \pi(w-z)},\quad w\in \bold Z\setminus D; \quad \chi(w)=\int_{ D}
{dA\over\pi( w-z)},$$
 then Eq.(3.6) can be written as
$$ {\partial \chi(z,t)\over\partial t}-C{\partial \chi(z,t)\over\partial x}=
\sum_{i=1}^N {q_i\over \pi (z-z_i)}.\eqno(3.7)$$
It is a first
order p.d.e. that is readily solved explicitly. The solution has
the form
$$ \chi(z,t)=\chi_0(z+Ct)+\int_0^t \sum_{i=1}^N {q_i d\tau \over \pi
(z+C(t-\tau)-z_i)}.
\eqno(3.8)$$
For the dipole, $ \chi_0(z)={A/z}, $ and the
integrand becomes $ \mu/ [\pi (z+C(t-\tau))^2],$ and therefore
$$ \chi(z,t)= {\mu\over Cz}+{AC-\mu\over C(z+Ct)}. $$
The first term is the Cauchy transform for a circle of area
$A_0=\mu/C$ centered at the origin; the second term corresponds to
the circle of the area $A-A_0$ centered at $z=-Ct$. It means that
at large $t$ the solution is combination of a steady-state circle
of area $A_0$ at the origin corresponding to the limiting
steady-state solution, and the  circle of the area $A-A_0$
``sinking" in the gravity field.

As general solution (3.8) is valid for any function $h$, it is
tempting to state, that an arbitrary initial domain of
sufficiently large area under the combined action of gravity and a
dipole at the origin $(z=0)$  eventually splits into two parts,
namely, a stationary disk of area $A_0=\mu/C$ centered at $z=0$,
and a ``sinking'' domain with the Cauchy transform of the form
$$ \chi_1=\chi_0(z+tC)-\mu/[C(z+tC)]. \eqno(3.9) $$
At $t=0$ the shape of the ``sinking'' domain is specified by the Cauchy transform
$$ h_1(z,0)=h_0(z)-\mu/(Cz). $$
It can be derived from the initial domain in the absence of gravity by placing
a sink at the origin and sucking the amount of fluid corresponding to the
equilibrium domain area $A_0$. If we now allow this new domain to slide
far
enough in the gravity field,  and then inject back the same amount of fluid
at the origin, we will get exactly the Cauchy transform specified by the
Eq.(3.9). Now we see, that this conjectured form of evolution will
indeed
occur, if the initial domain will evolve smoothly during the initial sucking of
fluid. It will be certainly so, if the initial domain itself can be
produced from a simply connected domain by injecting the amount of
fluid $A_0$ without violating the simply-connectedness condition.

This argument allows one to develop a number of explicit solutions
for domains evolving in gravity field in the presence of a dipole
at the origin.


Now we are going to consider some less trivial examples of equilibrium domains.
 It is worth noting that such domains can exist only for special combinations
of hydrodynamic sources and the external potential. Indeed,
if we let in the moment equation (2.6) $U=F(z)$, it results in
$$ \int_D |\omega(z)|^2dA =-\sum_{j=1}^N q_jF(z_j) $$
and since the l.h.s. of this equation is positive, the (complex)
electric potential should satisfy the condition
$$ -\sum_{j=1}^N q_jF(z_j)>0.\eqno (3.10) $$

 {\it A priori} the sum in the r.h.s. can be any complex number,
but for the equilibrium shape to exist, the r.h.s. should be real
and positive. Obviously, this inequality can hold only for special
form of potential. Say, in the case of a doublet source-sink of
equal strength both of them should lie on the same force line of
the electric field (${\rm Im} F(z_1)={\rm Im} F(z_2)$).

As we will see later, if some electric field sources ("charges")
are within the flow domain $D$, there is a continuous spectrum of
the equilibrium domain areas. For example, in the simplest case of
absence of hydrodynamic sources, $q_j=0$, there is no flow within
the equilibrium domain, the potential $\Phi={ const} \quad {\rm
in}\quad D,$ and boundary condition  (1.12) implies that $G={
const}\ $ along $\partial D$, so the  boundary should be a level
curve of the electric potential. Say, in the case of a single
electric source the equilibrium domains are circles of arbitrary
radius centered at the source. Notice, that  since the boundary of
the equilibrium domain in the absence of hydrodynamic sources is
just a level curve of electric potential the analytic continuation
 procedure reduces to the reflection principle of electrostatics applied in the $\zeta$-plane.
\vskip 12pt
{\it Example 1.} Let we have two hydrodynamic sources of the strengths
$q_1=-q_2=q$
at $z_1=a>0$, $z_2=b$ respectively, and an electric ``charge'' $Q$ at $z=0$.
Then
$$F(z)={Q\over 2\pi} \ln z,\eqno (3.11)$$
Inequality (3.10) implies that $qQ\ln(b/a)>0$, so that $b>a$ for positive $Q$.

Using reduction to the ``gravity'' case technique, we have in the potential
 plane $\tilde z= {(Q/2\pi)} \ln z$ the Cauchy transform
for the transformed domain $\tilde D=F(D)$
$$ h_{\tilde D}= {q\over \pi} \ln \left( {\tilde z -{Q\over 2\pi} \ln a\over
\tilde z-{Q\over 2\pi}\ln b}\right).\eqno (3.12)$$ Then the
conformal map $\tilde f$ of the unit disk $K$ on $\tilde D$ is
given by the expression (cf. \cite{EEK1})
$$\tilde f(\zeta)={q\over \pi} \ln {1+\alpha \zeta \over  1-\alpha \zeta}+ {Q\over 2\pi}
\ln \sqrt{ab}, \eqno (3.13) $$
with $\alpha$ determined from the equation
$$  {q\over \pi} \ln {1+\alpha^2 \over  1-\alpha^2}= {Q\over 2\pi}
\ln \sqrt{b\over a},\quad 0<\alpha <1.\eqno (3.14) $$
Therefore,
$$ \alpha=\sqrt{{(b/a)^{\lambda/2}-1\over (b/a)^{\lambda/2}+1}};
\quad \lambda ={Q\over 2q},$$
and
$$ f(\zeta)=e^{{2\pi\over Q}\tilde f(\zeta)}=\sqrt{ab}
\left({1+\alpha \zeta\over 1-\alpha\zeta}\right)^{1/\lambda}.
\eqno(3.15)$$
The solution is a circle for $\lambda =1$, i.e. $Q=2q$.
The solution remains physically sensible only for not too large
$b/a$; otherwise overlapping of different parts of predicted domain
occurs.
\vskip 6pt
\begin{small}
   Consider condition of non-overlapping of the mapping (3.15).
The condition of overlapping is
$$ f(\zeta)=f(\overline \zeta)= f(1/\zeta);\quad \zeta=e^{i\phi} $$
In our case it implies
 $$\left(1+\alpha\zeta \over 1- \alpha\zeta \right)^{1/\lambda}=
\left(1+\alpha\zeta^{-1} \over 1- \alpha\zeta^{-1} \right)^{1/\lambda}$$

In the plane
$$\zeta_1= {1+\alpha\zeta \over 1- \alpha\zeta} $$
the boundary is a circle of the radius $r_1=2\alpha/(1-\alpha^2)$
centered at $(1+\alpha^2)/(1-\alpha^2)$.  As
$$ z=C\zeta_1^{1/\lambda},$$
the overlapping occurs at the point of maximum $\arg(\zeta_1)$.  But
$$ \beta=\max(\arg(\zeta_1))=\sin^{-1} (2\alpha/(1+\alpha^2)).$$
So overlapping occurs at
$$ \beta=\pi\lambda;\quad 2\alpha/(1+\alpha^2)=\sin \pi\lambda $$
or
$$ \alpha=\tan{\pi Q\over 4 q}; \quad
{(b/a)^{\lambda/2}-1\over (b/a)^{\lambda/2}+1}=\tan^2{\pi Q\over 4 q}$$
$$ (a/b)=\left( \cos {\pi Q\over 2q} \right)^{4q/Q}. \eqno (3.16) $$

Therefore, non-overlapping equilibrium domain exists in the range of
parameters
$$ 1\ge a/b \ge (\cos \pi \lambda)^{(2/\lambda)}; \quad \lambda=Q/(2q).$$

For small $\lambda$ non-overlapping equilibrium domains exist only
in  a narrow range of $b/a$ close to unity.  If we let the ratio
$b/a$ tend to the critical value $ (\cos \pi \lambda)^{2/\lambda},
$ the area of the equilibrium domain increases rapidly, and the
domain acquires horseshoe shape.
For $\lambda=1/2$
the equilibrium domain remains simply connected for any $a/b$.
In the limiting case
$$f(\zeta)=\left (  {1-\zeta\over 1+\zeta}\right)^2 $$
the critical domain is the entire $z$ plane with a cut along the negative
real axis.
\end{small}
\vskip 6pt An example is presented in Fig.{\ref{Fig1}.
 \begin{figure}[tbp]
\begin{center}
\includegraphics[width=11cm]{EE01_1FF.epsc}\\
a.\\
 \includegraphics[width=12cm]{EE01_1FFB.epsc}\\
b.\\
 \caption{ Equilibrium domains. Flow is driven by a source at
$z=a$, and sink at $z=b$, an electric charge $Q$ is located
at $z=0$; plots   correspond to
q=1; a=1; b=4;
and
$Q=  0.2734, 0.2959, 0.3189, 0.3424, 0.3664, 0.3909$.
for curves 1-6 respectively (a). Figure (b)
 shows blow-up of the upper figure
illustrating that the electric charge is outside the
equilibrium domain.
}\label{Fig1}
 \end{center}
\end{figure}
For curve 1 the ratio $Q/2q$ is quite close to the critical value.
Consider now a limiting case when the hydrodynamic source and sink
collide, and form a dipole of the moment $\mu$. Formally, it
corresponds to
$$ b/a=e^\epsilon; \quad q={\mu\over \epsilon a},\quad \epsilon\rightarrow
0.
$$
Then Eq.(3.15) becomes
$$ f_\epsilon(\zeta)=a e^{\epsilon/2}\left({1+\alpha_\epsilon\zeta\over
     1-\alpha_\epsilon\zeta } \right)^{1/\lambda_\epsilon},
\quad \lambda_\epsilon={Q\over 2q}=
{\epsilon a Q\over 2\mu} .\eqno (3.17) $$
$$\alpha_\epsilon=\left({e^{\epsilon^2 aQ/4\mu}-1\over
e^{\epsilon^2 aQ/4\mu}+1}\right)^{1/2}\approx \epsilon
\sqrt{aQ\over 8\mu}. \eqno (3.18)$$ 
So we find
$$ f_0(\zeta)=\lim_{\epsilon \rightarrow 0} f_\epsilon(\zeta)=
\lim_{\epsilon \rightarrow 0} a
\left( 1+\epsilon \sqrt{aQ\over
2\mu}\zeta \right)^{2\mu \over \epsilon a Q}
=a \exp {\left( \zeta\sqrt{2\mu \over aQ}\right)  }. \eqno (3.19)$$
 This mapping corresponds to a
non-overlapping domain iff
$$\sqrt {2\mu / aQ}\le \pi;\quad {\mu / aQ}\le \pi^2/2.\eqno (3.20)$$
At greater values of ${\mu / aQ}$,  there is no simply connected equilibrium
 domain. It can be conjectured that in this case the electric field is too weak to
prevent breakthrough caused by the hydrodynamic dipole.

Figure \ref{Fig2}
shows shapes of equilibrium domain for $\mu=1, a=1$;
$Q$= 0.2026;  0.2410;
    0.2866;    0.3408;    0.4053.
 \begin{figure}[tbp]
\begin{center}
\includegraphics[width=12cm]{EE01_2FF.epsc}\\
a.\\
 \includegraphics[width=12cm]{EE01_2FFB.epsc}\\
 b.\\
 \caption{
 Equilibrium domains. Flow is driven by a dipole
of moment $\mu$ at $z=a$, an electric charge $Q$ is located
at $z=0$; plots   correspond to
$\mu$=1; a=1;
and
$Q$= 0.2026;  0.2410;
    0.2866;    0.3408;    0.4053.
for curves 1-5 respectively (a). Figure (b)
 shows blow-up of the upper figure
illustrating that the electric charge is outside the
equilibrium domain.
}\label{Fig2}
 \end{center}
\end{figure}
\subsection{Singular points technique.}

{\it Example 2.}  Consider now the equilibrium domains
corresponding to a hydrodynamic dipole of the moment $\mu$ and an
electric source of the strength $Q$ both located at $z=0$. In this
case, the differential $d[F(f(\zeta))]$ has singularities only at
0 and  $\infty$. The general method described above implies
 that
$$ d[F(f(\zeta))]=\left( {P\over \zeta}+R\right) d\zeta,  \eqno (3.21) $$
$$ F(f(\zeta)) =P\ln \zeta +R\zeta + C, \eqno (3.22) $$
$$ f(\zeta)=Ae^{2\pi F/Q}=A\zeta^{2\pi P/Q}e^{(2\pi R/Q)\zeta}.
\eqno (3.23) $$ Since $f(\zeta)$ is a conformal map, $f'(0)\ne
0,\infty$. Therefore,
$$ P={Q\over 2\pi}; \quad  f(\zeta)=A\zeta e^{B\zeta}.
\eqno (3.24) $$ In order to find $B$, we use the the moment
relation
$$\int_D\overline{\omega(z)}U'(z)dS=\mu U'(0).\eqno (3.25) $$
In our case it assumes the form
$$\int_D{Q\over 2\pi \overline{z}}U'(z)dS=\mu U'(0).\eqno (3.26) $$
Choosing $U=z$, we get
$$\int_D{dS\over \overline{z}}={2\pi \mu\over Q}.\eqno (3.27) $$
Substituting
$$z=A\zeta e^{B\zeta}, $$
we get
$$\int_K {|A+AB\zeta|^2|e^{B\zeta}|^2\over
\overline \zeta Ae^{B\overline\zeta}} d\sigma=
$$
$$
\int_K {(A+AB\zeta)(A+AB\overline\zeta)e^{B\zeta}\over A\overline
\zeta } d\sigma ={2\pi \mu\over Q}.\eqno (3.28) $$ Evaluating the
integral in the l.h.s. of Eq.(3.28) we find
$$ B={2\mu\over QA};\quad f(\zeta) = A\zeta e^{2\mu\zeta\over QA}.
\eqno (3.29) $$ The parameter $A$ still indeterminate is a size
parameter. It should satisfy the condition that $f(\zeta)$ is
single-valued. Therefore critical values of the parameter
correspond to

(1) $f'(\zeta)=0,\quad \zeta\in \partial K$, or

(2) $f(\zeta)=f(\overline \zeta)=f( \zeta^{-1}),\quad \zeta \in \partial K$.

The first condition results in:
 $$f'(\zeta)=A(1+B\zeta)e^{B\zeta}=0,\
\zeta\in \partial K, \Rightarrow B=1,
\ {\rm or} \ {2\mu \over AQ}=1. \eqno (3.30) $$

The second condition implies
 $$ 1=f(\zeta)/f(\zeta^{-1})
=\zeta^2e^{{2\mu\over QA}(\zeta-\zeta^{-1})}=
 e^{2i\phi +2i{2\mu\over QA}\sin\phi}.\eqno (3.31) $$
Then the critical condition becomes
$$ L(\phi)={\phi +{2\mu\over QA}\sin\phi}=\pi k,\quad 0< \phi < \pi.
\eqno (3.32) $$ As $L(\pi)=\pi$,  a root in the segment $(0,\pi)$
appears as derivative
$$L'(\pi)=1+{2\mu\over QA}\cos\pi=1-{2\mu\over QA}$$
becomes negative, or at
$${2\mu \over AQ}=1.\eqno (3.33) $$
This condition coincides with (3.30). Hence a simply connected
equilibrium domain exists for
$${2\mu \over AQ}\le 1,\quad A\ge {2\mu\over Q} .\eqno (3.34) $$
Thus there exists a continuous spectrum of sizes of equilibrium domains that
is bounded from below.

  This result can be interpreted as inability of a given ``charge"
to prevent breakup due to action of hydrodynamic dipole, if the domain
area is too small, or the domain boundary is too close to the dipole.

Figure \ref{Fig3} shows boundaries of the equilibrium domains
desribed mapping of the unit disk given by Eq.(3.29) for
$Q=1$, $\mu=1$, $A=1, \sqrt{2}, 2,  2\sqrt{2}$, and 4
 for curves 1-5 respectively.
 \begin{figure}[tbp]
\begin{center}
\includegraphics[width=12cm]{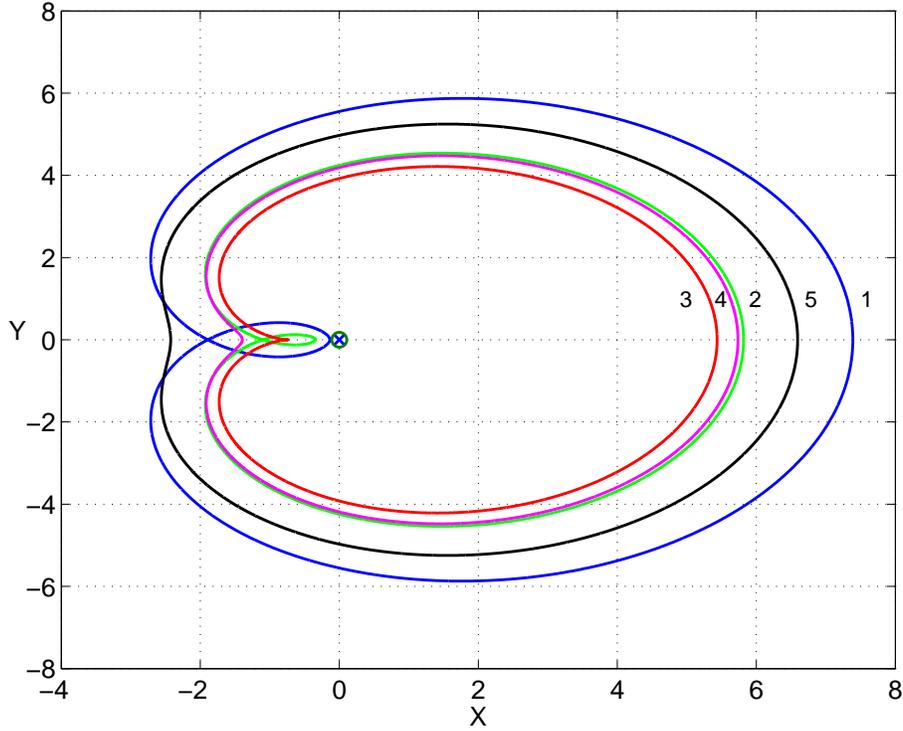}\\
 \caption{
 Equilibrium domains. Flow is driven by a dipole
of moment $\mu$ and an electric charge $Q$ located
at $z=0$; plots   correspond to
$\mu$=1; Q=1;
and
$A=1, \sqrt{2}, 2,  2\sqrt{2}$, and 4
 for curves 1-5 respectively. Self-intersecting boundaries
1 and 2 correspond to non-physical domains.
}\label{Fig3}
 \end{center}
\end{figure}
{\it Example 3.} Now we consider interaction of a hydrodynamic
quadrupole at the origin with two  ``charges" of strength $Q$ at
$z=\pm a$ outside the equilibrium domain $D$. We fix the conformal
mapping of $K$ on $D$ by conditions $f(0)=0; f'(0)>0$. In this
case $F(f(\zeta))$ has a pole of second order at infinity, and
hence
$$ F(f(\zeta))=-\alpha \zeta^2.\eqno(3.35)$$
and
$$ F(z)={Q\over 2\pi} \ln\left(1-{z^2\over a^2}\right).\eqno(3.36)$$
Therefore
$$ f(\zeta)=z=a\sqrt{1-e^{-{2\pi \alpha\over Q}\zeta^2}},\quad \alpha>0
.\eqno(3.37)$$ The parameter $\alpha$ depends on the strength
$\beta$ of the hydrodynamic quadrupole, namely, the pole
coefficient of $F(f(\zeta))$ at $\infty$ is the same as that of
$h(f(\zeta))$ at $\zeta=0$. As
$$ h={\beta\over 2\pi z^2} ,\eqno(3.38)$$
the requirement implies
$${\beta\over 2\pi a^2 {2\pi \alpha\over Q}}=\alpha;\quad
\alpha^2={\beta Q\over 4\pi^2a^2} .$$ Therefore the solution
exists for $\beta>0$, and
$$\alpha={\sqrt{\beta Q}\over 2\pi a},$$
Thus,
$$ f(\zeta)=a\sqrt{1-
\exp{\left(-\sqrt {\beta\over a^2Q}\zeta^2\right)} } =a\zeta
\sqrt{1- \exp{\left(-\sqrt {\beta\over
a^2Q}\zeta^2\right)}\over\zeta^2}. \eqno(3.39)$$ This mapping
remains univalent until
$$ \exp{\left(-\sqrt {\beta\over a^2Q}\zeta\right)}$$
remains single-valued, so that
$${1\over a}\sqrt {\beta\over Q}\le \pi; \quad {\beta\over a^2Q}\le \pi^2.
\eqno(3.40)$$
Figure \ref{Fig4} shows mapping of the unit disk
given by Eq.(3.39) for $\beta=1, Q=1, a=0.2251, 0.2677,
0.3183=1/\pi, 0.3785$ (curves 1-4 respectively).
 \begin{figure}[tbp]
\begin{center}
\includegraphics[width=12cm]{EE01_4FF.epsc}\\
a.\\
 \includegraphics[width=12cm]{EE01_4FFB.epsc}\\
 b.\\
 \caption{
 Equilibrium domains. Flow is driven by a quadrupole
of moment $\beta$ and two electric charges $Q$ located
at $z=\pm a$; plots   correspond to
$\beta$=1; Q=1;
and
$A=0.2677;\    0.3183=1/\pi;\    0.3785;\    0.4502
$
 for curves 1-4 respectively (a). Blowup of Fig.\ref{Fig4},a.
The charge remains outside the equilibrium domain; only non-intersecting
boundaries correspond to physically admissible equilibrium domains.
}\label{Fig4}
 \end{center}
\end{figure}

\section{ Non-harmonic External Field. Reduction to the Riemann-Hilbert problem}
The problem of finding stationary shapes of flow domain in an
external field can be approached in a different way that allows
extension to non-harmonic external field potential of special
form.


Let $D$ be an equilibrium domain for a specified set of
hydrodynamic sources and multipoles corresponding to logarithmic
singularities and poles of the velocity potential $W(z)$ in the
field of the external force potential $G(x,y)$.

Then $W(z)$ is an analytic (while may be multivalued) function in
$D$ having prescribed set of singularities; its differential
$dW(z)$ is a meromorphic function in $D$ and
 $$ W(z)|_{z\in \partial D}= \tilde G(z,\overline z).\eqno(4.1)$$
Here,
$$\tilde G(z,\overline z)\equiv G\left({1\over 2}(z+\overline z), {1\over 2i}
(z-\overline z)\right).\eqno(4.2)$$

 We are going to show that in number of cases due to special form of the
  potential $G(x,y)$ it proves to be possible to find the domain $D$
explicitly. Consider once more the conformal map $f: K\rightarrow
D$ and define
$$ \Theta(\zeta)=W(f(\zeta)),\quad \zeta\in K. \eqno(4.3)$$
This function is analytic up to singularities of the specified
type (poles and logarithmic singular points) in $K$ and assumes
real values along the boundary $\partial K$. Then it can be
analytically continued into the entire complex plane $\zeta$ using
the symmetry principle:
$$ \Theta(\zeta)=\overline{\Theta\left({1\over\overline \zeta}\right)},\quad |\zeta|>1.
  \eqno(4.4)$$
Then
$$\Theta(\zeta)=\sum_{j=1}^N{q_j\over 2\pi}
[\ln(\zeta-\zeta_j)+ \overline {\ln(\zeta - \overline
\zeta_j^{-1})}].\eqno(4.5)$$ Hence $\Theta(\zeta)$ is known in the
entire complex plane $\zeta$ up to locations of the singularities
$\zeta_j$. At the boundary of the unit disk
$$ \Theta(\zeta)= \tilde G(z,\overline z).\eqno(4.6)$$
For given $\zeta_j$, it is an equation for the conformal mapping
$f(\zeta)$ that can be written as

$$ \tilde G(f(\zeta),f^*(1/\zeta))=\Theta(\zeta).\eqno (4.7)$$

In general, it is not clear how to determine the conformal map
$f(\zeta)$ from this equation. However, it proves to be possible
under some additional assumptions on $G$.

Some of these particular cases are presented below.

\subsection {Harmonic velocity potential}

Suppose that $G$ is a harmonic function with maybe a finite set of
logarithmic singular points within $D$, and let $F(z)$ be
respective complex potential. Then
$$ G(z,\overline z)={1\over 2} (F(z)+\overline {F(z)})=
{1\over 2}( F(z)+ F^*(z)).\eqno(4.8)$$
 Introducing this expression
into Eq.(4.7), we get
$$ F^*(f^*(1/\zeta))=2\Theta(\zeta)- F(f(\zeta)).\eqno(4.9)$$
Therefore $ F^*(f^*(1/\zeta))$ has finitely many singularities
within $K$. Those inside $K$ correspond to singularities of
$\Theta(\zeta)$ and $F(f(\zeta))$ inside $K$ [i.e both
``hydrodynamic'' and ``electric''
 singularities], those outside $K$ are explicitly determined by the singularities of $\chi(\zeta)=F(f(\zeta))$.
Therefore the derivative
$$ {d\over d\zeta}F(f(\zeta))$$ is meromorphic in the entire  $\zeta$ plane, and hence it is rational with
 number and order of singular points known beforehand.
This allows one to write down its explicit expression up to a
number of indetermined coefficients. Then the conformal mapping is
expressed as
$$ f(\zeta)=F^{-1}(\chi(\zeta)),\eqno (4.10)$$
and it remains to write and solve a set of equations for location
and strength of singularities. It is the case considered
prevoiusly.

\subsection{ Unidirectional external field}
Suppose now that
$$G=H(x)=H({1\over 2}(z+\overline z)),\quad H'(x)>0.\eqno (4.11)$$
It means, that the external ``force'' has only $x$-component that
is independent of $y$. Then
$$\Theta(\zeta)=H({1\over 2}(f(\zeta)+f^*(1/\zeta))), \eqno (4.12)$$
$$f(\zeta)+f^*(1/\zeta)=2H^{-1}(\Theta(\zeta)),\quad \zeta\in \partial K.
\eqno (4.13)$$ The functions $f(\zeta)$ and $f^*(1/\zeta)$ are
analytic respectively inside and outside the unit circle. It is
the Riemann-Hilbert problem that is solved using the Cauchy-type
integral (cf.\cite{Muskh1},\cite{Gakhov}):

$$ f(\zeta)=
{1\over \pi i}\oint_{\partial K} {H^{-1}(\Theta(u))\over u-\zeta}
du -{1\over 2\pi i}\oint_{\partial K} {H^{-1}(\Theta(u))\over u}
du. \eqno (4.14)$$

{\it Example 4.1}. Let $H(x)=x^2$, and the flow is generated by a
dipole at a location $z=x_0>0$. Then
$$ W(z)\sim {\mu\over z-x_0},\quad z\rightarrow x_0.
\eqno (4.15)$$ We assume that $x_0=f(0)$. Then $\Theta(\zeta)$
should have poles at $\zeta=0$ and $\zeta=\infty$, and, therefore,
$$ \Theta(\zeta)=\alpha\left(\zeta+{1\over\zeta}\right)+\beta.\eqno (E.6)$$
Introducing these expressions into Eq.(4.14), we get
$$ f(\zeta)=
{1\over \pi i}\oint_{\partial K}
{\sqrt{\alpha(u+u^{-1})+\beta}\over u-\zeta} du -{1\over 2\pi
i}\oint_{\partial K} {\sqrt{\alpha(u+u^{-1})+\beta}\over u} du.
\eqno (4.17)$$ Then the dipole location is given by the
expression:
$$x_0={1\over 2\pi i}
\oint_{\partial K} {\sqrt{\alpha(u+u^{-1})+\beta}\over u} du,
\eqno (4.18)$$ while for its strength $\mu$ we find
$$ {\mu \over f'(0)}= \alpha;\quad f'(0)=
{\mu\over\alpha}= {1\over \pi i}\oint_{\partial K}
{\sqrt{\alpha(u+u^{-1})+\beta}\over u^2} du . \eqno (4.19)$$
Equations (4.18) and (4.19) serve to find $\alpha$ and $\beta$ for
given $x_0$ and $\mu$. They can be reduced to equations
$$x_0={1\over\pi }\int_0^\pi
\sqrt{2\alpha\cos\varphi+\beta}d\varphi; \quad {\mu\over\alpha}=
{2\over \pi }\int_0^\pi \sqrt{2\alpha
\cos\varphi+\beta}\cos\varphi d\varphi. \eqno (4.20)$$ Relations
Eq.(4.20) are shown in Fig.\ref{Fig5},a.
\begin{figure}[tbp]
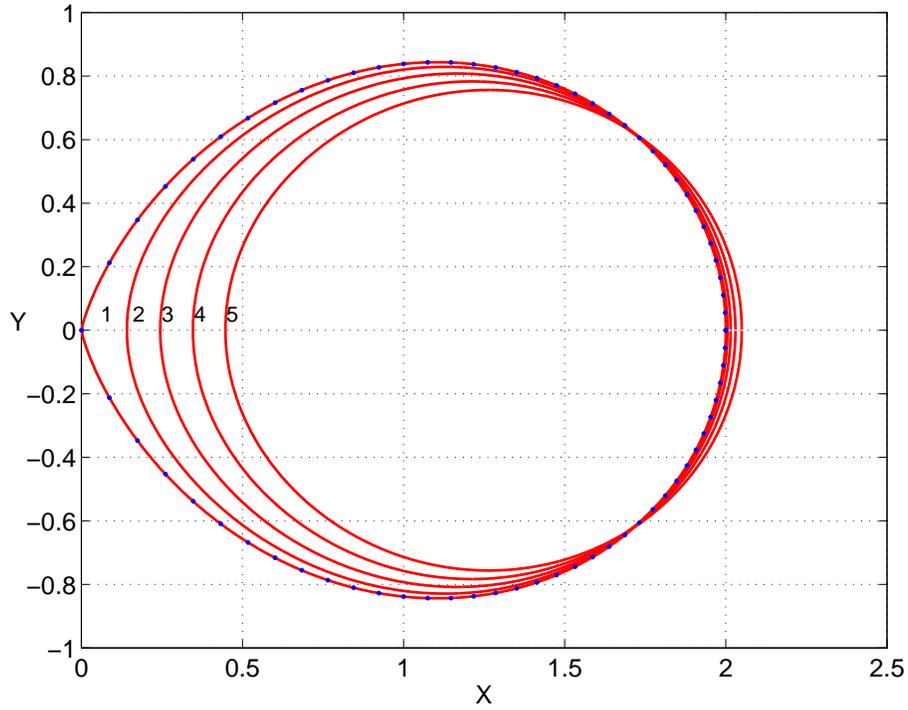

\begin{center}
\includegraphics[width=12cm]{EE01_5FF.epsc}\\
 a.\\
 \includegraphics[width=12cm]{EE01_6FF.epsc}\\
 b.
\caption{ {\textbf a}:  Relation between geometric parameters of
the equilibrium domain and relative strength of the
dipole; {\textbf b}:  Shape of equilibrium domains for $B= 2.00;
2.02;
 2.06;    2.12;    2.20$
(curves 1-5 respectively).}
 \label{Fig5}
 \end{center}
\end{figure}
Using these relations, we can construct explicitly the equilibrium
domains predicted by conformal mapping Eq.(4.18). Some results are
presented in Fig.\ref{Fig5},b.

\subsection{ Axially-symmetric  external field}

We assume now that the external potential has the form
$$ G=H(x^2+y^2)=H(z\overline z),\quad H'(r)>0, r\ne 0 {\rm\  in} \ D,\
r=(z\overline z)^{1/2}.
 \eqno (4.21)$$
that corresponds to an a radially-symmetric external field with
the symmetry axis outside the equilibrium domain $D$. Equation
(4.4) implies
$$ f(\zeta) f^*(1/\zeta)= H^{-1}(\Theta(\zeta)), \eqno (4.22)$$
or
$$ \ln f(\zeta) + \ln f^*(1/\zeta)= \ln H^{-1}(\Theta(\zeta)). \eqno (4.23)$$
As by assumption $f(\zeta) \ne 0;\quad \zeta \in \overline D$, the
logarithms in the l.h.s. of this equation are analytic functions
respectively in the unit disk and outside it, and therefore we
once more have the Riemann-Hilbert problem. Its solution is
$$ f(\zeta)=\exp\left(
{1\over 2\pi i}\oint_{\partial K}{\ln H^{-1}(\Theta(u))\over
u-\zeta} du - {1\over 4\pi i}\oint_{\partial K}{\ln
H^{-1}(\Theta(u))\over u} du\right).
 \eqno (4.24)$$
These expressions allow us to restore the shape of the equilibrium
domain provided the expression for $\Theta(\zeta)$ can be guessed
using properties of the hydrodynamic singularities.

{\it Example 4.2}. Let $H(x^2+y^2)=r^2$, and the flow is generated
by a dipole at a location $z=r_0$. Then
$$ W(z)\sim {\mu\over (z-r_0)},\quad z\rightarrow r_0.
\eqno (4.25)$$ We assume that $r_0=f(0)$. Then repeating argument
of the previous subsection, we find the same expression (4.16) for
$\Theta(\zeta)$, and keeping in mind that in our case
$H^{-1}(X)=X$, we have,
 upon introducing this expression into Eq.(4.24),
$$ f(\zeta)=\exp\left(
{1\over 2\pi i}\oint_{\partial K}{\ln (\Theta(u))\over u-\zeta} du
- {1\over 4\pi i}\oint_{\partial K}{\ln (\Theta(u))\over u}
du\right)
$$
$$
={\exp \left({1\over 2\pi i}\oint_{\partial K}{(\ln [\alpha
(u+1/u)+\beta]\over u-\zeta} du\right) \over \exp\left({1\over
4\pi i}\oint_{\partial K}{(\ln [\alpha (u+1/u)+\beta]\over u} du
\right)}.
 \eqno (4.26)$$
Characteristic shapes of the equilibrium domains predicted by the
mapping Eq.(4.26)are shown in Fig.\ref{Fig7}.
\begin{figure}[tbp]
\begin{center}
\includegraphics[width=12cm]{EE01_7FF.epsc}\\
a.\\
 \includegraphics[width=12cm]{EE01_7FFB.epsc}\\
b.\\
\caption{Equilibrium domains for flow driven by a dipole in
quadratic axisymmetric external potential field for $B= 2.0; 2.021
2.061;    2.121;    2.201$ (curves 1-5 respectively), (a); blow-up
of Fig.\ref{Fig7},a, (b).}
 \label{Fig7}
 \end{center}
\end{figure}

\subsection{ External field depending on a harmonic function
} Let the external potential depend on a function harmonic up to
specified logarithmic singularities,
$$ G=H(T(x,y)), $$
$$\Delta T=\sum_{m=1}^M Q_m\delta(x-x'_m,y-y'_m).
 \eqno (4.27)$$
Then $T(x,y)$ is  the real part of an analytic function $\Xi(z)$
having specified logarithmic singularities, and
$$ G=H({1\over 2}(\Xi(z)+\Xi^*(\overline z))).
\eqno (4.28)$$ Let function $H^{-1}$ be rational and all
``hydrodynamic singularities'' correspond to multipoles (there is
no logarithmic singularities corresponding to sources). Then from
Eq.(4.28)

$$\Xi(f(\zeta))+\Xi^*(f^*(1/\zeta))= 2H^{-1}(\Theta(\zeta)),\quad \zeta\in \partial K, \eqno (4.29)$$
or, denoting
$$Z(\zeta)= \Xi(f(\zeta)),$$
$$Z(\zeta))+Z^*(1/\zeta))=
 H^{-1}(\Theta(\zeta)),\quad \zeta\in \partial K.
\eqno (4.30)$$ It is essentially the same equation as Eq.(4.13),
and it can be solved using the same technique. Then the conformal
mapping
$$f(\zeta)=\Xi^{-1}(Z(\zeta)).\eqno (4.31)$$
\vskip 12pt
{\it Example 4.3.} Let $$T=x^2-y^2;\quad H(T)
=\sqrt{x^2-y^2};\quad \Xi(z)={1\over 2}z^2;\eqno (4.32)$$ and let
the flow is generated by a single dipole of the strength $\mu$ at
$z=a>0$,
$$ W(z)\sim {\mu\over z-a},\quad z\rightarrow a.\eqno (4.33)
 $$
Then $\Theta(\zeta)$ is expressed by Eq.(4.16), and
$$Z(\zeta)=
{1\over \pi i}\oint_{\partial K}
{\sqrt{\alpha(u+u^{-1})+\beta}\over u-\zeta} du -{1\over 2\pi
i}\oint_{\partial K} {\sqrt{\alpha(u+u^{-1})+\beta}\over u} du,
\eqno (4.34)$$
$$ f(\zeta)= \sqrt{2Z}.\eqno(4.35)$$
Then for $\alpha$ and $\beta$ we have equations
$$ a=f(0)=\sqrt {{2\sqrt{\alpha}\over\pi}\int_0^\pi\sqrt
{2\cos\varphi +\beta/\alpha}d\varphi};\eqno (4.36)$$
$$ {\mu\over f'(0)}= \alpha;\quad f'(0)=
{\mu\over\alpha}= (2/Z(0))^{1/2}Z'(0)=$$
$$
{4\sqrt{\alpha}\over a\pi}\int_0^\pi \sqrt{2\cos\varphi
+\beta/\alpha}\cos\varphi d\varphi.\eqno (4.37)$$ Shapes of the
equilibrium domains for flow driven by a dipole at $z=1$ in the
external field  corresponding to Eq.(4.32) are shown in
Fig.\ref{Fig8}.
\begin{figure}[tbp]
\begin{center}
\includegraphics[width=12cm]{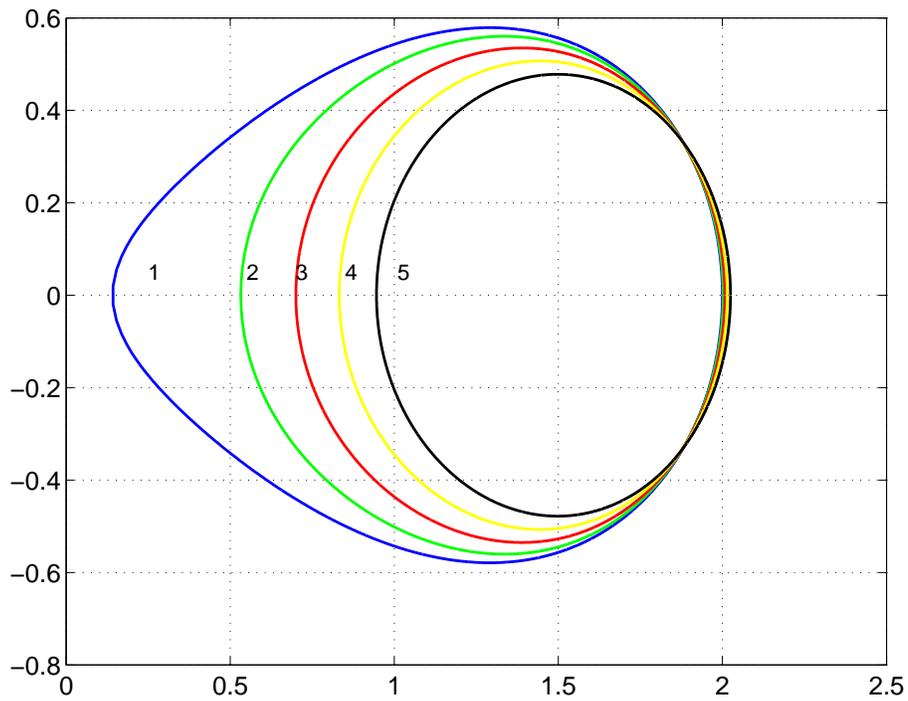}
\caption{Equilibrium domains for flow driven by a dipole in
external potential field of the form Eq.(4.32)
 for $B= 2.0001;\    2.0201;\    2.0601;\    2.1201;\
    2.2001$
(curves 1-5 respectively).}
 \label{Fig8}
 \end{center}
\end{figure}

\subsection{Non-planar Hele-Shaw cell}

Consider a non-planar Hele-Shaw cell in constant gravity field.
Let $(x,y)$ be coordinates in the horizontal plane, and $h(x,y)$
is elevation of a cell point over the horizontal plane. Then
assuming $h=h(x)$ it is possible to introduce the conformal
coordinate
$$ z=s(x)+iy,\quad s(x)=\int_0^x\sqrt{1+h'(t)^2}dt,\eqno (4.38)$$
 so that the problem reduces to that of planar Hele-Shaw cell with effective
potential
$$ H(z)=h(x)=h(s^{-1}({\rm Re} z)).\eqno (4.39)$$

Similarly, if
$$ h(x,y)=K(r);\quad r=\sqrt{x^2+y^2},\eqno (4.40)$$
then the conformal coordinate is
$$ z=e^{i\varphi}R(r);\quad R(r)=\exp \int_1^r\sqrt{1+K'(\rho)^2}{d\rho\over \rho},
 \eqno (4.41)$$
and the effective potential is
$$ H(z,\overline z)=h(R^{-1}(|z|)).\eqno (4.42)$$

\section{Possible applications}

In this Section we show that the mathematical model considered
above can be used to model electroosmotically-driven flow in a
thin gap between to infinite parallel walls provided that the gap
is filled with two immiscible fluids having equal electric
conductivities, viscosity of the fluid in the exterior of the
domain $D(t)$ is negligible, and electroosmotic coefficients of
the two fluids are different. Therefore, there exists at least one
non-trivial physical situation corresponding to the mathematical
problem considered in this paper.

\subsection{Electrokinetic Effect: Physics, Available data }

Electrokinetic effect consists in generation of electric current
by fluid flow through porous media or thin gaps between solid
walls, and in the reverse effect of inducing flow by application
of electric field. It is the last case, usually referred as {\it
electroosmosys} that serves as primary motivation of presented
theory. Electrokinetic phenomena are caused by  difference in
mobility of ions, some of which are fixed at the surface of the
solid skeleton (matrix) of the porous medium, or the solid walls,
while dissolved counterions can move with the fluid within the gap
or porespace, or force it to move,
 if an electric field is applied.

Macroscopically, the flow and electric current are governed by the
equations
$$ {\bf u }=-{k\over \eta}(\nabla p -\xi \nabla \psi), \eqno( 5.1)$$
$$ {\bf I }=-S(\nabla \psi -C\nabla p),\quad C=1/\xi. \eqno( 5.2)$$
Here, $k$ is the medium permeability, $\eta$ is the fluid
viscosity, $S$ is the fluid electric conductivity $C$ is the
electrokinetic coupling coefficient (see
\cite{Overbeek},\cite{Dukhin}, \cite{EOsmADL} for details).

 Both ``streaming potentials", i.e. electric fields
generated by fluid flow,
 and ``electroosmotic flow", the flow driven by electric
potential differential, have important applications.
Electroosmotic flow is used in soil remediation and prevention of
moisture penetration in underground structures. Recently,
electroosmotic flow is also actively studied as an element of
microfluidic devices, when flow in narrow gaps or channels is
driven by electric potential \cite{EOsm}. Presumably, it is this
class of flows, to which
 the presented
above theory can find some applications.

Namely, we consider flow driven both by pressure gradient and
external electric field in a narrow plane gap between two solid
non-conducting walls. We assume, that due to significant fluid
conductivity the flow effect on electric current is negligible. In
this case, Eqs.(5.1) and (5.2) become
$$ {\bf u }=-{k\over \eta}(\nabla p -\xi \nabla \psi), \eqno( 5.3)$$
$$ {\bf I }=-S\nabla \psi. \eqno( 5.4)$$
here, ${\bf u}(x,y)$ and ${\bf I}(x,y)$ are averaged over the gap
thickness flow velocity and electric current; they satisfy the
continuity equations (conservation laws)
$$ \nabla {\bf\cdot u }= q_u(x,y);\quad
\nabla {\bf\cdot I }= Q_I(x,y);\eqno(5.5)$$ $k=h^2/12$; pressure
and the electric potential are functions of the in-plane
coordinates $(x,y)$.

Now we assume that the gap is filled by two fluids, one of them,
within time-dependent plane domain $D(t)$, is characterized by the
viscosity $\eta$, conductivity $S$ and electrokinetic coupling
coefficients $C$ and $\xi$; another, outside $D(t)$, is filled by
another fluid with viscosity $\eta_1$ and conductivity $S_1$, and
electrokinetic coupling coefficients $C_1$ and $\xi_1=1/C_1$. Then
at the boundary $\Gamma(t)=\partial D(t)$ we have
$$ p^+=p^-;\quad \psi^+=\psi^-;\quad
  u^+_n=u_n^-;\quad I_n^+=I_n^-.\eqno(5.6)$$
For given densities of the volume ($q_u$) and electric ($q_I$)
sources Eqs.(5.3)-(5.6) define a free boundary problem of coupled
pressure/electroosmotically driven flow in the gap.

Now we consider a particular case when both fluids have the same
conductivity, $S_1=S$. Then $\psi(x,y)$ satisfies the equation
$$ \Delta \psi(x,y)= {q_I\over S}\eqno (5.7)$$
in the entire plane.

It can be considered as known.
 Let now the viscosity of the external fluid outside
 $D(t)$ be negligible.
Then, assuming that there is no net flux to infinity, we have
$$ p^+(x,y)=\xi_1\psi(x,y,t)+const_1, \quad (x,y)
\in \bold Z\setminus D(t).\eqno (5.8)$$ Now we define ``effective
pressure" as
$$ P(x,y)=p(x,y)-\xi_1\psi(x,y)-const_1.\eqno(5.9)$$
Then we have
$$ {\bf \nabla\cdot\ u}=0,\quad
 {\bf u}=-{k\over\eta}(\nabla P-(\xi-\xi_1)\nabla \psi);
\quad {\bf x}\in D(t); \quad P({\bf x})=0,   \quad {\bf x}\in
\partial D(t) \eqno(5.10)$$ It is, up to notations, the problem
considered in this paper. Above examples show that external
electric field can be used to confine the flow to a finite domain
$D$. 

\end{document}